\documentclass[pdflatex,iicol,sn-mathphys-num]{sn-jnl}
\usepackage{graphicx,amsmath,amssymb,booktabs,bm}
\usepackage[T1]{fontenc}

\newcommand{\st}{s_\theta}
\newcommand{\ct}{c_\theta}
\begin{document}
\interdisplaylinepenalty=10000
\title[Thermal contact in Majorana dark matter]{Thermal contact and direct detection in a broken $Z_4$ Majorana dark matter model}
\author[1]{\fnm{XinXin} \sur{Qi}}\email{qxx@dlut.edu.cn}
\author*[1]{\fnm{Hao} \sur{Sun}}\email{haosun@dlut.edu.cn}
\affil[1]{\orgdiv{Institute of Theoretical Physics, School of Physics},\orgname{Dalian University of Technology},\orgaddress{\street{No. 2 Linggong Road},\city{Dalian},\postcode{116024},\state{Liaoning},\country{China}}}
\abstract{
We consider a broken $Z_4$ Majorana dark matter model in which the fermion mass is generated by a real singlet scalar. For small scalar mixing, freeze-out is dominated by $\chi\chi\to h_2h_2$, whereas momentum exchange with the Standard Model bath becomes inefficient. In this region, the usual assumption $T_\chi=T$ is not always adequate. We therefore evolve the dark matter abundance and kinetic temperature simultaneously. Although $T_\chi$ is still close to the bath temperature around chemical freeze-out, the later departure from kinetic equilibrium reduces the residual $p$-wave annihilation rate and leaves a larger relic abundance. In the sub-TeV region considered here, the difference between the two-moment and single-temperature calculations reaches $37.1\%$ at fixed parameters. Refitting the Yukawa coupling to reproduce $\Omega_\chi h^2=0.12$ increases the spin-independent scattering cross section by up to $18.6\%$. A momentum-resolved Fokker--Planck calculation gives a similar shift, $17.7\%$, at a representative point. The LZ upper limit on the scalar mixing changes by less than $0.3\%$, since kinetic equilibrium is already well maintained in that part of parameter space. The lower mixing boundary obtained from the comparison of the two relic calculations therefore indicates where the common-temperature approximation loses accuracy; it is not an experimental lower bound. These results apply to the specified low-reheating broken-phase cosmology and to the region where threshold and strong nonrelativistic-potential effects remain small.}
\maketitle

\section{Introduction}

Secluded dark matter models provide a simple way of separating the process that fixes the relic abundance from the interaction that connects the dark sector to the Standard Model (SM) \cite{Pospelov:2007mp}. In singlet-fermion Higgs-portal models \cite{Kim:2008pp}, this separation is controlled by the scalar mixing. The annihilation rate can remain large even when the coupling to the SM becomes small, so chemical freeze-out and kinetic decoupling need not occur at the same temperature. Once this happens, the usual assumption that dark matter shares a common temperature with the thermal bath may no longer be adequate. The impact of kinetic decoupling on thermal freeze-out has been studied using coupled moment equations and momentum-dependent treatments \cite{Binder:2017rgn,Binder:2021bmg}.

We study this effect in a real-singlet model with one Majorana fermion and a spontaneously broken $Z_4$ symmetry. After symmetry breaking, a residual $Z_2$ stabilizes the fermion, while the singlet vacuum expectation value generates its mass. Related singlet models with discrete symmetries have been discussed in Refs.~\cite{Yaguna:2023kyu,Yaguna:2024jor,Qi:2025jpm,Qi:2025znq}. In the parameter region considered here, with small scalar mixing and $m_2<m_\chi$, the dominant annihilation channel is $\chi\chi\to h_2h_2$. The relic abundance is therefore mainly controlled by interactions inside the dark sector, whereas kinetic contact with the SM plasma can become inefficient at a later stage of freeze-out.

To quantify this effect, we compare the usual single-temperature relic calculation with a coupled evolution of the dark matter abundance and kinetic temperature. Elastic scattering with both SM particles and the scalar bath is included. The thermal history of secluded sectors has been studied in several related settings. Ref.~\cite{Bringmann:2021decoupled} considered relic formation in thermally decoupled dark sectors, while Ref.~\cite{Duan:2024thermal} followed the number densities and temperatures of a Majorana fermion and a light scalar, including finite-transfer elastic effects. Ref.~\cite{DiMauro:2025jsb} studied several minimal secluded dark matter models, including a singlet-generated Dirac fermion mass. Here we concentrate on the broken-$Z_4$ Majorana case and on how kinetic decoupling changes the relic-density fit and the associated direct-detection prediction.

We perform independent single-temperature and two-moment relic fits over a grid in dark matter mass, scalar-to-dark-matter mass ratio, and scalar mixing. The resulting accuracy boundary is then compared directly with the LZ limit \cite{LZ:2024zvo}. At selected points we also solve a momentum-resolved evolution equation to check the two-moment treatment. The numerical analysis is restricted to a low-reheating broken-phase cosmology and to parameter regions sufficiently far from the scalar-pair threshold and from sizeable nonrelativistic potential effects.

The paper is organized as follows. Section~\ref{sec:model} introduces the model and the quantities relevant for annihilation and direct detection. The cosmological setup and the parameter region used in the numerical analysis are described in Sec.~\ref{sec:history}. In Sec.~\ref{sec:moments} we present the coupled evolution equations for the dark matter abundance and kinetic temperature. The numerical results are given in Sec.~\ref{sec:results}. We summarize our conclusions in Sec.~\ref{con}.

\section{Model description}\label{sec:model}

We extend the SM by a real singlet scalar $S$ and a two-component
fermion $\chi$. Under the $Z_4$ symmetry,
\begin{equation}
S\to -S,
\qquad
\chi\to i\chi .
\end{equation}
The relevant terms in the Lagrangian are
\begin{equation*}
\mathcal{L}
\supset
i\chi^\dagger\bar{\sigma}^{\mu}\partial_\mu\chi
+\frac{1}{2}(\partial_\mu S)^2
-\frac{1}{2}\left(yS\chi\chi+\mathrm{h.c.}\right)
-V ,
\label{eq:lagrangian}
\end{equation*}
with
\begin{equation}
V=
-\mu_H^2 H^\dagger H
-\frac{\mu_S^2}{2}S^2
+\lambda_H(H^\dagger H)^2
+\frac{\lambda_S}{4}S^4
+\frac{\lambda_{HS}}{2}S^2H^\dagger H .
\label{eq:potential}
\end{equation}
All parameters are taken to be real, with $y>0$.

After symmetry breaking, we write
\begin{equation}
H^0=\frac{v+h}{\sqrt{2}},
\qquad
S=w+s ,
\end{equation}
and choose $w>0$. The scalar mass eigenstates are
\begin{equation}
h_1=c_\theta h-s_\theta s,
\qquad
h_2=s_\theta h+c_\theta s ,
\label{eq:mixing}
\end{equation}
where $c_\theta\equiv\cos\theta$ and
$s_\theta\equiv\sin\theta$. We identify $h_1$ with the observed Higgs
boson and set $m_1=125~\mathrm{GeV}$, while $h_2$ denotes the
additional singlet-like scalar.

The singlet vacuum expectation value generates the Majorana mass
\begin{equation}
m_\chi=yw .
\label{eq:DMmass}
\end{equation}
The fermion terms after symmetry breaking are therefore
\begin{equation}
\mathcal{L}
\supset
-\frac{1}{2}m_\chi\chi\chi
-\frac{1}{2}y
\left(-s_\theta h_1+c_\theta h_2\right)\chi\chi
+\mathrm{h.c.}
\label{eq:fermion_interactions}
\end{equation}
The vacuum $w\neq0$ breaks $Z_4$ to the residual $Z_2$ under which
$\chi\to-\chi$, ensuring the stability of the fermion.

We choose $(m_\chi,m_2,y,s_\theta)$ as independent inputs, with
$w=m_\chi/y$ and $v=246~\mathrm{GeV}$. The parameters of the scalar
potential are then
\begin{equation*}
\lambda_H
=
\frac{m_1^2c_\theta^2+m_2^2s_\theta^2}{2v^2},
\quad
\lambda_S
=
\frac{m_1^2s_\theta^2+m_2^2c_\theta^2}{2w^2},
\end{equation*}
\begin{equation*}
\lambda_{HS}
=
\frac{(m_2^2-m_1^2)s_\theta c_\theta}{vw},
\end{equation*}
and
\begin{equation*}
\mu_H^2
=
\lambda_Hv^2+\frac{1}{2}\lambda_{HS}w^2,
\quad
\mu_S^2
=
\lambda_Sw^2+\frac{1}{2}\lambda_{HS}v^2 .
\end{equation*}

\subsection{Tree-level consistency}

The potential is bounded from below for
$\lambda_H,\lambda_S>0$ and
$\lambda_{HS}>-2\sqrt{\lambda_H\lambda_S}$.
The mixed vacuum additionally satisfies
$4\lambda_H\lambda_S>\lambda_{HS}^2$ and is the tree-level global
minimum, apart from the two degenerate choices $w=\pm|w|$.

Using normalized two-particle states, the scalar $J=0$ partial-wave
eigenvalues are
\begin{equation}
\frac{1}{16\pi}
\left\{
2\lambda_H,\,
\lambda_{HS},\,
\Lambda_\pm
\right\},
\end{equation}
where
\begin{equation*}
\Lambda_\pm
=
\frac{
6\lambda_H+3\lambda_S
\pm
\sqrt{(6\lambda_H-3\lambda_S)^2+4\lambda_{HS}^2}
}{2}.
\end{equation*}
The eigenvalues $2\lambda_H$ and $\lambda_{HS}$ are ninefold and
fourfold degenerate, respectively. Requiring
$|\operatorname{Re}a_J|<1/2$ gives
\begin{equation}
|2\lambda_H|<8\pi,
\qquad
|\lambda_{HS}|<8\pi,
\qquad
|\Lambda_\pm|<8\pi .
\end{equation}
The Yukawa interaction gives the additional high-energy constraint
\cite{Allwicher:2021rtd}
\begin{equation}
y^2<4\pi .
\end{equation}

\subsection{Annihilation and direct detection}

For $m_2<m_\chi$ and small scalar mixing, the relic abundance is
mainly controlled by
\begin{equation}
\chi\chi\to h_2h_2 .
\end{equation}
We evaluate the full tree-level amplitude, including the $t$- and
$u$-channel fermion diagrams, the two scalar $s$-channel contributions,
and their interference. The same model parameters are used in the
elastic-scattering calculation. The isolated $t/u$ contribution is
given in Appendix~\ref{app:ann}.

The spin-independent dark-matter--nucleon scattering cross section at
zero momentum transfer is
\begin{equation}
\sigma_N^{\mathrm{SI}}
=
\frac{\mu_{\chi N}^2}{\pi}
\left[
\frac{y s_\theta c_\theta\,m_N f_N}{v}
\left(
\frac{1}{m_1^2}-\frac{1}{m_2^2}
\right)
\right]^2 ,
\label{eq:DD}
\end{equation}
where
\begin{equation}
\mu_{\chi N}
=
\frac{m_\chi m_N}{m_\chi+m_N}
\end{equation}
is the dark-matter--nucleon reduced mass and $f_N$ is the effective
scalar nucleon form factor. We use the default scalar nucleon matrix
elements implemented in \textsc{micrOMEGAs} and evaluate the proton
and neutron cross sections separately.

For comparison with the isospin-conserving LZ limit, we define the
xenon-equivalent nucleon cross section
\begin{equation}\label{xeeq}
\sigma_{\mathrm{Xe,eq}}
=
\left[
\frac{
Z\sqrt{\sigma_p^{\mathrm{SI}}}
+
(A-Z)\sqrt{\sigma_n^{\mathrm{SI}}}
}{A}
\right]^2 ,
\end{equation}
with $Z=54$ and $A=131.293$. The proton and neutron scalar amplitudes
have the same sign in the parameter region considered here.
The quantity $\sigma_{\mathrm{Xe,eq}}$ is compared with the observed
LZ 90\% confidence-level limit at the corresponding dark matter mass.

\section{Specified cosmology and approximation domain}\label{sec:history}
Spontaneous breaking of an exact discrete symmetry may lead to the formation of domain walls. We therefore consider a cosmological history in which the observable Universe lies in a single $w>0$ vacuum patch before the final stage of inflation. The inflationary Hubble scale is taken to be smaller than the relevant scalar mass scales, suppressing transitions between the two degenerate vacua. We assume instantaneous reheating followed by radiation domination and entropy conservation, with
\begin{equation}
T_{\max}=T_R=\frac{m_\chi}{20}.
\label{eq:TR}
\end{equation}
This choice fixes the cosmological background used throughout the numerical analysis. Scenarios with prolonged reheating or $T_{\max}\gg T_R$ are not considered, since thermal restoration of the discrete symmetry could then reintroduce the domain-wall problem.

It remains to check that the scalar background is not significantly displaced at the reheating temperatures used here.
At leading order, the thermal corrections to the scalar mass parameters are
\begin{align}
c_H&=\frac{3g^2+g'^2}{16}+\frac{\lambda_H}{2}
+\frac{y_t^2}{4}+\frac{\lambda_{HS}}{24},\\
c_S&=\frac{\lambda_S}{4}+\frac{\lambda_{HS}}{6}
+\frac{y^2}{12}.
\label{eq:thermal}
\end{align}
For the singlet direction, the relevant zero-temperature curvature after minimizing over the Higgs background is
\begin{equation}
\mu_{S,\mathrm{eff}}^2
=
\mu_S^2-\frac{\lambda_{HS}\mu_H^2}{2\lambda_H}.
\end{equation}
Across the parameter region used below,
\begin{equation}
\frac{c_ST_R^2}{\mu_{S,\mathrm{eff}}^2}
=
5.5\times10^{-4}\text{--}1.64\times10^{-3},
\end{equation}
while the corresponding scale
$T_S^{\mathrm{LO}}=(\mu_S^2/c_S)^{1/2}$
lies between 462 and 1030 GeV. This is well above the reheating range
$T_R=15$--40 GeV adopted in the numerical analysis.

The low-temperature background was also evaluated with the regulated one-loop two-field potential described in Appendix~\ref{app:thermal}. At $T_R$, all retained reference points satisfy
$|h(T_R)/v-1|<0.0061$ and
$|s(T_R)/w-1|<10^{-6}$ without daisy resummation. With the Arnold--Espinosa prescription, the corresponding maxima are $0.0036$ and $3.5\times10^{-6}$. Using these background values in the tree-level parametrization changes
the single-temperature relic abundance by less than $10^{-4}$ in the
tested parameter region. The relic calculation is therefore performed in the broken-phase zero-temperature parametrization. The high-temperature restoration region is not reached in the cosmological history considered here.

\subsection{Weak-potential regime}\label{wrg}

For $m_2<m_\chi$, exchange of $h_1$ and $h_2$ gives an attractive
nonrelativistic potential,
\begin{equation}
V_\chi(r)
=
-\frac{\alpha_1 e^{-m_1 r}+\alpha_2 e^{-m_2 r}}{r},
\end{equation}
\begin{equation}
(\alpha_1,\alpha_2)
=
\frac{y^2}{4\pi}(\st^2,\ct^2).
\end{equation}
To avoid regions where potential effects would have to be included
explicitly in the relic calculation, we define
\begin{equation}
\zeta
=
m_\chi
\left(
\frac{\alpha_1}{m_1}
+
\frac{\alpha_2}{m_2}
\right).
\end{equation}
The bound derived in Appendix~\ref{app:potential} shows that
$\zeta<1$ excludes potential bound states.

We also solve the $p$-wave radial equation over
$10^{-3}\leq v_{\rm rel}\leq1$ and retain only points for which
\begin{equation}
\max S_1(v_{\rm rel})<1.10 .
\end{equation}
The $10\%$ cut is simply used to keep the analysis in a region where
Sommerfeld corrections remain small. For the points in the following discussion, $\zeta=0.043$--$0.127$ and
$S_1^{\rm max}=1.031$--$1.095$.
Further details are given in Appendix~\ref{app:potential}.

\section{Chemical and kinetic evolution}\label{sec:moments}
We use \textsc{micrOMEGAs}~7.1.1 together with matrix elements generated by
\textsc{CalcHEP} \cite{Alguero:2023zol,Belyaev:2012qa}. The dark matter
abundance is written as
\begin{equation}
Y=\frac{n_\chi}{s_{\rm ent}},
\end{equation}
where $n_\chi$ is the dark matter number density and $s_{\rm ent}$ is the
entropy density. The dark matter kinetic temperature is described by
$q=T_\chi/T$, with $T$ the temperature of the thermal bath. The same
temperature-dependent effective degrees of freedom are used in the
thermodynamic quantities and in the rates exported from
\textsc{micrOMEGAs}.

We evolve the first two nonrelativistic moments of the Boltzmann equation,
assuming a Maxwell--Boltzmann form for the dark matter momentum
distribution. The second variable is chosen as
\begin{equation}
Q=Yq,
\end{equation}
which remains well defined when the initial dark matter abundance vanishes.
It is convenient to introduce
\begin{equation}
x\equiv\frac{m_\chi}{T},
z\equiv\ln x,
C=1+\frac{1}{3}\frac{d\ln h_{\rm eff}}{d\ln T},
A=\frac{C s_{\rm ent}}{H}.
\end{equation}
where $h_{\rm eff}(T)$ denotes the effective number of entropy degrees of
freedom and $H$ is the Hubble expansion rate. We further define
\begin{equation}
a(T)=\langle\sigma v\rangle_T,
\qquad
k(T)=\frac{d\ln a(T)}{d\ln T},
\end{equation}
where $a(T)$ is the thermally averaged annihilation rate and $k(T)$ measures
its logarithmic temperature dependence. The evolution equations are then
\begin{align}
\frac{dY}{dz}
&=
-A\left[a(T_\chi)Y^2-a(T)Y_{\rm eq}^2\right],
\label{eq:moment1}
\\
\frac{dQ}{dz}
&=
(1-2C)Q
+2C\frac{\gamma}{H}(Y-Q)-A[
(1+\frac{k(T_\chi)}{3})
\nonumber\\
&\quad
a(T_\chi)YQ
-
(1+\frac{k(T)}{3})a(T)Y_{\rm eq}^2
].
\label{eq:moment2}
\end{align}
Here $Y_{\rm eq}$ is the equilibrium yield and $\gamma(T)$ is the
momentum-transfer rate. The factors involving $k(T)$ account for the
energy carried away by annihilating particles and are relevant for the
$p$-wave process considered here. Their derivation is given in
Appendix~\ref{app:moments}. We denote the coupled solution by 2M, while
the calculation obtained by imposing $T_\chi=T$ is referred to as 1T.

The momentum-transfer rate is calculated from the elastic-scattering
amplitudes \cite{Bringmann:2006mu},
\begin{align}
\gamma(T)
&=
\sum_i
\frac{g_i}{6m_\chi T}
\int\frac{d^3k}{(2\pi)^3}
f_i(1\!\pm\!f_i)
\frac{|\bm{k}|}{E_i}
\nonumber\\
&\quad\times
\int_{-4p_{\rm cm}^2}^{0}
dt\,(-t)\frac{d\sigma_{\chi i}}{dt}.
\label{eq:gamma}
\end{align}
Here $i$ labels the particles in the thermal bath, $g_i$ is their
number of internal degrees of freedom, $f_i$ is the equilibrium
distribution function, and $E_i$ is the corresponding particle
energy. The upper (lower) sign in $1\pm f_i$ applies to bosons
(fermions). The quantity $p_{\rm cm}$ is the center-of-mass momentum
of the $\chi i$ system, and $t$ is the Mandelstam variable. The sum
includes SM fermions and antifermions, $W$, $Z$, $h_1$ and $h_2$,
with their finite masses retained.

The scalar mediator remains in thermal equilibrium over the temperature
range relevant for freeze-out. We check this using
\begin{equation}
\frac{\Gamma_2 K_1(m_2/T)}
     {H K_2(m_2/T)},
\end{equation}
where $\Gamma_2$ is the total decay width of $h_2$ and $K_1$ and $K_2$
are modified Bessel functions of the second kind.
Its minimum value at reheating is $6.76\times10^7$ over the retained
parameter region. Thus the scalar bath may keep $T_\chi$ close to $T$ near chemical freeze-out even when direct momentum exchange between dark matter and SM particles becomes inefficient at later times.

The evolution starts at $x_R=m_\chi/T_R=20$ and is followed to
$x=1000$. We have checked that the final abundance is insensitive to
the initial dark matter density at the reheating temperatures used in
the analysis.

\section{Numerical results and experimental comparison}\label{sec:results}
\subsection{Grid and kinetic accuracy}
We first determine $y$ from $\Omega_{1T}h^2=0.12$ at
$\st=10^{-3}$, using the Planck relic abundance as reference
\cite{Planck:2018vyg}. We restrict the analysis to the sub-TeV
parameter region satisfying the open-channel, weak-potential,
thermal-background and tree-level consistency conditions discussed in
Sec.~\ref{sec:history}.

\begin{figure*}[t]
\centering\includegraphics[width=.96\textwidth]{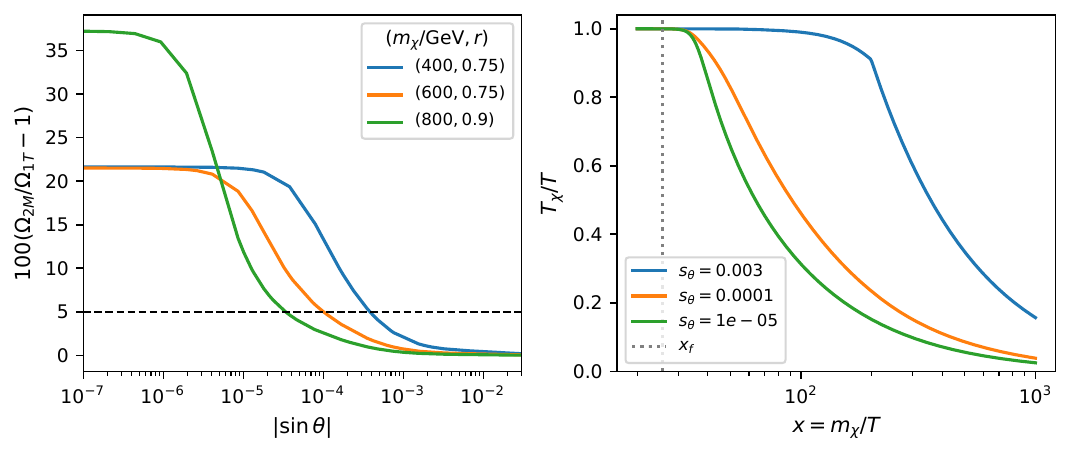}
\caption{Left: change in the relic abundance on the fixed single-temperature reference surface when kinetic evolution is included. The horizontal line denotes a five-percent approximation criterion. Right: the temperature ratio for the 400 GeV, $r=0.75$ point. Its proximity to unity at $x_f$ does not prevent later kinetic decoupling from modifying residual annihilation. The displayed curves are two-moment results; selected points are independently checked with a momentum-resolved elastic-FP calculation.}\label{fig:kinetics}
\end{figure*}

The effect of kinetic decoupling is quantified by
\begin{equation}
\delta_{\rm kin}
=
\frac{\Omega_{2M}}{\Omega_{1T}}-1,
\qquad
\delta_{\rm kin}(\st^{5\%})=0.05 .
\label{eq:criterion}
\end{equation}
For $m_\chi=400$~GeV and $r=0.75$ with $r\equiv m_2/m_\chi$, the correction increases from
about $0.8\%$ at $\st=0.003$ to $21.4\%$ at $\st=10^{-5}$.
As shown in Fig.~\ref{fig:kinetics}, $T_\chi$ remains close to $T$
around chemical freeze-out and separates only later, while residual
$p$-wave annihilation is still active.  Here $x_f$ denotes the chemical freeze-out point used in the reference 1T calculation.
The common-temperature
approximation can therefore receive a sizeable correction even when
$T_\chi/T\simeq1$ at freeze-out.

The five-percent boundary moves to smaller mixing as the dark matter
mass increases. For
$(m_\chi,r)=(400~{\rm GeV},0.75)$,
$(600~{\rm GeV},0.75)$ and
$(800~{\rm GeV},0.9)$, we find
\begin{equation}
\st^{5\%}
=
3.82\times10^{-4},
1.00\times10^{-4},
3.36\times10^{-5},
\end{equation}
respectively.

\subsection{Direct detection and electroweak precision}
For spin-independent scattering, we compare the model prediction with the observed power-constrained 90\% confidence limit from the 4.2 tonne-year LZ search \cite{LZ:2024zvo}. Using Eqs.\eqref{eq:DD} and \eqref{xeeq} , we determine the corresponding upper limit on the scalar mixing for each reference slice.

\begin{figure*}[t]
\centering\includegraphics[width=.96\textwidth]{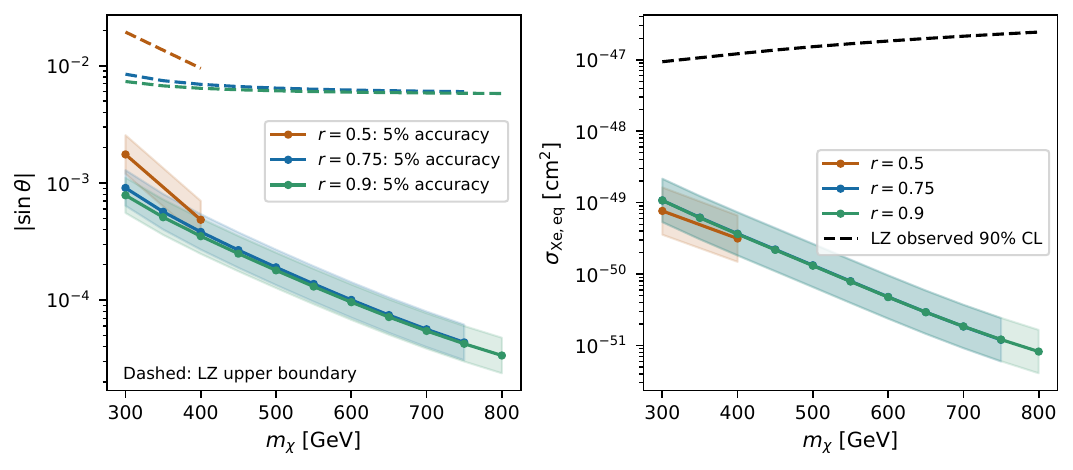}
\caption{Mixing boundaries (left) and their xenon-equivalent SI cross sections (right). Solid curves impose a five-percent agreement between two-moment and single-temperature abundances. Dashed curves give the observed LZ upper bound. Shading varies the total elastic rate by factors of $1/2$ and 2; it is not a confidence interval. Only retained grid points are connected. The solid lower curves express a calculation-accuracy requirement and do not exclude the region below them.}\label{fig:window}
\end{figure*}

Figure~\ref{fig:window} compares the LZ upper limit with the five-percent accuracy boundary, $\st^{5\%}$. An interval between the two is present for all retained slices. The lower boundary should not be interpreted as an experimental constraint: it only marks the point below which the single-temperature treatment differs from the two-moment result by more than five percent. Smaller mixing angles remain viable once the coupled evolution is used and the Yukawa coupling is refitted to the relic density.

The sensitivity of this boundary to the momentum-transfer rate is estimated by varying $\gamma$ by a factor of two. For $m_\chi=400$~GeV and $r=0.75$, $\st^{5\%}$ changes from
$3.82\times10^{-4}$ to $2.69\times10^{-4}$ and
$5.41\times10^{-4}$ for the two variations. The resulting range is shown as the shaded band in Fig.~\ref{fig:window}. 

Constraints from Higgs data are evaluated with \textsc{HiggsTools}~1.3.0 \cite{Bahl:2022igd}, using the scalar widths and rescaled SM couplings of the model. All retained points, including those at the LZ boundary, satisfy the HiggsSignals and HiggsBounds tests used in the analysis. We also include the one-loop scalar contributions to the oblique parameters $(S,T,U)$ following Ref.~\cite{Grimus:2008nb}. In the relevant small-mixing region these corrections are very small, with typical values of $S$ and $T$ at the LZ boundary of order $10^{-6}$. Electroweak precision constraints therefore do not further restrict the interval shown in Fig.~\ref{fig:window}.

\subsection{Relic refits and kinetic checks }\label{sec:refit}
To assess the impact on direct detection, we refit the Yukawa coupling
at fixed $(m_\chi,r,\st)$ so that the two-moment calculation reproduces
$\Omega_\chi h^2=0.12$. For the three representative slices
$(m_\chi,r)=(400,0.75)$, $(600,0.75)$ and $(800,0.9)$, the required
coupling at $\st=10^{-5}$ increases by $5.40\%$, $4.57\%$ and
$3.06\%$ relative to the corresponding 1T fits. The resulting
spin-independent cross sections are larger by approximately
$11.1\%$, $9.4\%$ and $6.2\%$, respectively. The refitted
small-mixing solutions remain well below the LZ limit, showing that
the five-percent accuracy boundary is not a lower experimental bound.

We also compared the two-moment treatment with more detailed kinetic
calculations at representative points. Using the finite-transfer scalar
collision term changes the final abundance by less than $0.04\%$ in the
tested region. A momentum-resolved calculation gives final
abundances within about $1\%$ of the two-moment result at small mixing
and moves the five-percent accuracy boundary by only a few percent.
Further details of these checks are given in Appendix~\ref{app:checks}.

\subsection{Parameter-space map}\label{sec:parammap}
We extend the analysis to a broader sub-TeV region in
$(m_\chi,r,s_\theta)$, performing independent 1T and 2M relic fits
at each parameter point. Along the 2M-refitted surface, we define
\begin{equation}
\delta_\Omega
=
\frac{\Omega_{\rm 2M}h^2}{\Omega_{\rm 1T}h^2}-1 ,
\label{eq:deltaOmega}
\end{equation}
where both abundances are evaluated at the same parameters obtained
from the 2M relic fit. The corresponding five-percent boundary,
$s_{5\%}^{2M}$, is defined by
\begin{equation}
\delta_\Omega\left(s_{5\%}^{2M}\right)=0.05 .
\label{eq:refit5}
\end{equation}
The direct-detection predictions are evaluated separately using the
Yukawa couplings obtained from the independent 1T and 2M relic fits.

\begin{figure*}[t]
\centering\includegraphics[width=.97\textwidth]{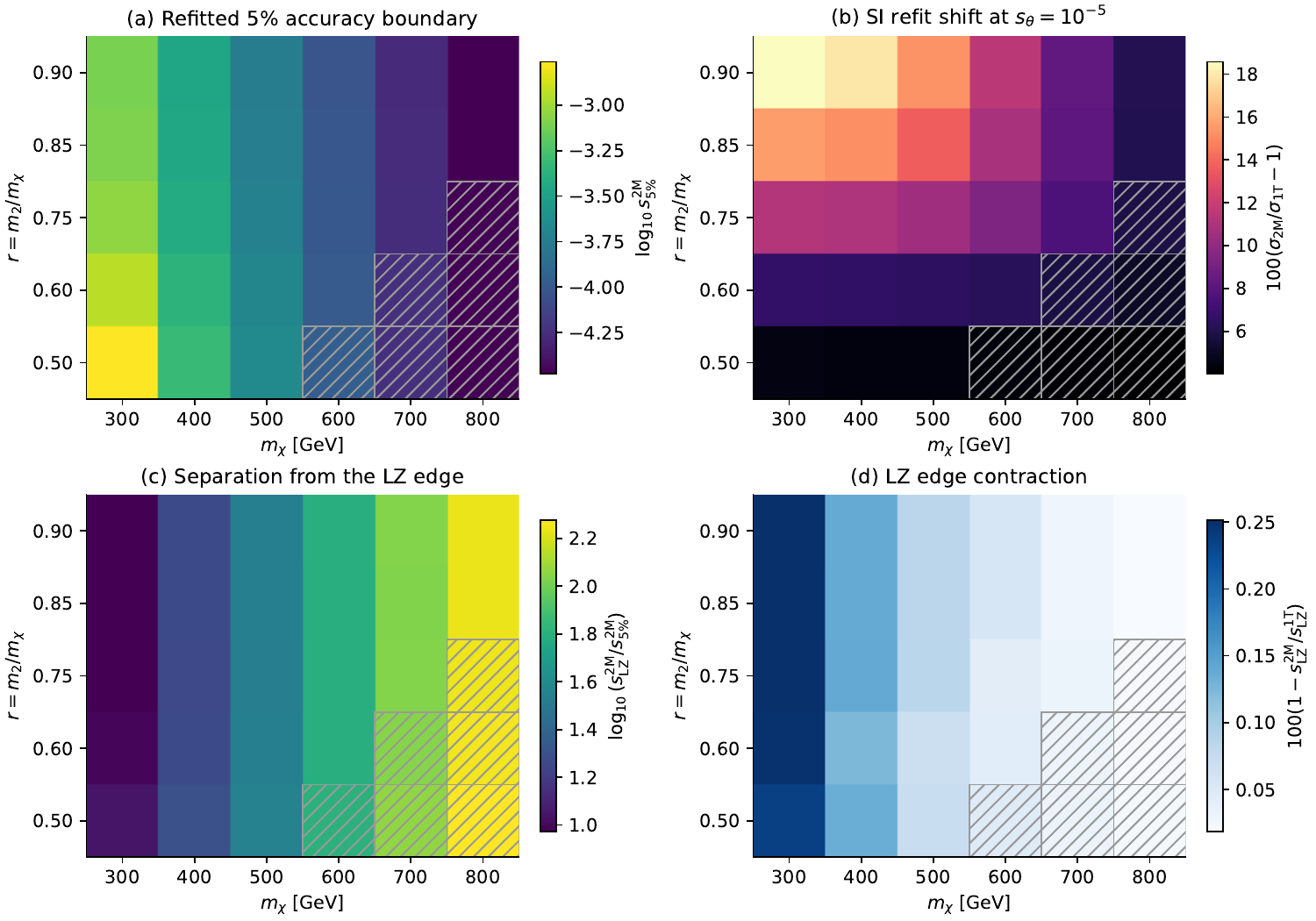}
\caption{Discrete mass--ratio map with independent 1T and 2M relic refits. (a) Five-percent boundary on the 2M-refitted surface. (b) Change of the SI prediction at $s_\theta=10^{-5}$ after refitting. (c) Separation between that accuracy boundary and the 2M LZ intersection. (d) Contraction of the LZ mixing edge. Each colored cell represents a computed mass--ratio pair, not a continuously interpolated exclusion. Hatching marks cells in which the adopted study-domain conditions are not satisfied over the sampled mixing range. The $r>0.9$ stress tests are omitted from this figure.}\label{fig:parammap}
\end{figure*}

Figure~\ref{fig:parammap} summarizes the results in the region satisfying
the theoretical and cosmological conditions discussed above. The
fixed-input relic correction reaches $37.1\%$, and refitting the relic
abundance increases the spin-independent cross section by up to
$18.6\%$. The largest effect occurs at small scalar mixing. Near the
LZ upper boundary, however, kinetic contact remains efficient and the
shift in the mixing limit is only $0.019\%$--$0.251\%$. The lower
boundary in Fig.~\ref{fig:parammap} therefore marks the loss of accuracy
of the common-temperature relic calculation rather than an experimental
lower limit.

We further examine the thermal-shape approximation with a
momentum-resolved Fokker--Planck (FP) calculation. The annihilation
term uses the full invariant matrix element for
$\chi\chi\to h_2h_2$, while elastic scattering is treated with the
relativistic FP operator. 
The FP calculation follows the same trend as the two-moment result. We denote the corresponding boundary obtained from the
momentum-resolved calculation by $s_{5\%}^{\rm FP}$.
At the point where the largest relic correction is found, the FP and
2M abundances differ by $1.35\%$, compared with the $37.1\%$
difference between 2M and 1T. Recomputing the five-percent criterion
with the momentum-resolved evolution shifts the mixing boundary by
$2.3\%$--$3.6\%$. We also repeat the relic-density fit using the FP
abundance. For $(m_\chi,r)=(300~\mathrm{GeV},0.9)$, the resulting
increase in the SI cross section is $17.7\%$, compared with $18.6\%$
from the 2M calculation; for $(400~\mathrm{GeV},0.6)$ the
corresponding values are $6.36\%$ and $6.60\%$. The momentum-resolved
calculation therefore gives essentially the same relic-refit effect as
the two-moment treatment in the tested region.

\begin{table}[t]
\caption{Direct FP checks of the five-percent accuracy boundary along the 2M-refitted surface. The last row is outside the threshold-margin domain. The two mixing columns are multiplied by $10^4$.}\label{tab:mapFP}
\centering\small
\begin{tabular}{rrrrr}\hline
$m_\chi$ [GeV] & $r$ & $10^4s_{5\%}^{\rm 2M}$ & $10^4s_{5\%}^{\rm FP}$ & shift [\%]\\\hline
300 & 0.50 & 17.324 & 16.742 & $-3.36$\\
300 & 0.90 & 7.773 & 7.593 & $-2.32$\\
400 & 0.60 & 4.289 & 4.171 & $-2.74$\\
800 & 0.90 & 0.3343 & 0.3221 & $-3.63$\\
400 & 0.98 & 3.089 & 3.007 & $-2.64$\\\hline
\end{tabular}
\end{table}
Larger corrections appear close to the scalar-pair threshold. For
$(m_\chi,r,\st)=(400~\mathrm{GeV},0.98,10^{-5})$, the relic-density
shift is $49.1\%$ in the 2M calculation and $46.9\%$ in the
momentum-resolved FP treatment. Refitting the relic abundance with the
FP result increases the SI cross section by $23.2\%$ relative to the
1T fit. We do not include this point in the main quantitative results,
since thermal pole-mass effects and threshold corrections are not
included in the present calculation.

Finite-transfer corrections remain much smaller: at the tested edge
points they change the final abundance by less than
$6.0\times10^{-5}$ in relative terms. The larger effect near threshold
therefore comes from the increased sensitivity of the annihilation
process itself, rather than from the treatment of scalar momentum
transfer.

\section{Conclusions}\label{con}
The broken-$Z_4$ Majorana model provides a simple setting in which
scalar-pair annihilation can remain efficient while kinetic contact
with the Standard Model becomes weak. In the sub-TeV region considered
here, the dark matter temperature can depart from the bath temperature
after chemical freeze-out and still modify the final relic abundance.
Thus, the condition $T_\chi\simeq T$ at freeze-out is not sufficient to
establish the accuracy of a single-temperature relic calculation.

Along the 2M-refitted grid, evaluating the 1T abundance at the same input parameters gives a maximum difference of $37.1\%$.
Refitting the Yukawa coupling to
$\Omega_\chi h^2=0.12$ increases the predicted spin-independent
scattering cross section by up to $18.6\%$. The same behaviour is found
in the momentum-resolved calculation, which gives a $17.7\%$ increase
at the point with the largest effect considered in the main analysis.
The corresponding five-percent accuracy boundaries move by only a few
percent, while the LZ upper mixing limit changes much less. The lower
mixing boundary should therefore be interpreted as the point at which
the common-temperature approximation loses accuracy, rather than as an
experimental lower limit on the scalar mixing.

These results refer to the low-reheating broken-phase cosmology adopted
in this work and to the parameter region away from the scalar-pair
threshold and sizeable nonrelativistic potential effects. The relic
calculation is performed at tree level, with finite-transfer and
momentum-resolved calculations used to check the kinetic treatment in
representative regions. A more complete treatment would be required
close to threshold or in the strong-potential regime.

\backmatter
\section*{Acknowledgements}
Hao Sun is supported by the National Natural Science Foundation of China under Grants No. 12075043 and No. 12147205. XinXin Qi is supported by the National Natural Science Foundation of China under Grant No. 12447162. ChatGPT 6.0 (Astra) was used for code
debugging, numerical cross-checks, literature comparison, and language
editing. The authors checked the code changes and numerical results and take
responsibility for the scientific conclusions.

\begin{appendices}
\section{Annihilation cross-section check}\label{app:ann}
For the isolated $t/u$ diagrams with a CP-even coupling $g=y\ct$, the nonrelativistic expansion is $\sigma v=b_{tu}v_{\rm rel}^2+O(v_{\rm rel}^4)$, with
\begin{equation}
b_{tu}=\frac{y^4\ct^4}{24\pi m_\chi^2}
\frac{\sqrt{1-r^2}(9-8r^2+2r^4)}{(2-r^2)^4}.
\end{equation}
Here $r=m_2/m_\chi<1$. The light-mediator limit is $3y^4\ct^4/(128\pi m_\chi^2)$.  The scalar $s$-channel diagrams interfere with this contribution
and are retained in the numerical relic calculation. For $h_i\to\chi\chi$, we define
$g_{\chi\chi 1}=-y s_\theta,
g_{\chi\chi 2}=y c_\theta$.
When kinematically allowed, the corresponding decay width is\begin{equation}
\Gamma(h_i\to\chi\chi)=\frac{g_{\chi\chi i}^2m_i}{16\pi}
\left(1-\frac{4m_\chi^2}{m_i^2}\right)^{3/2}
\end{equation}
which provides a check of the identical-Majorana normalization.

\section{Nonrelativistic annihilation energy moment}\label{app:moments}
For a Maxwell--Boltzmann distribution, the mean kinetic energy of an
annihilating pair follows from the temperature dependence of the
thermal average $a(T)$,
\begin{equation}
\left\langle E_{\rm kin,1}+E_{\rm kin,2}\right\rangle_{\rm ann}
=
3T+T^2\frac{d\ln a}{dT}.
\end{equation}
With
\begin{equation}
k(T)=\frac{d\ln a}{d\ln T},
\end{equation}
the kinetic energy removed per annihilated particle is
$(3+k)T/2$. Relative to the usual nonrelativistic energy moment
$3T/2$, this gives the factor $1+k/3$ appearing in
Eq.~\eqref{eq:moment2}. For a $p$-wave rate, $a(T)\propto T$ and
hence $k=1$, so the corresponding annihilation weight is $4/3$.
The inverse process is evaluated at the bath temperature by detailed
balance.

The remaining terms in Eq.~\eqref{eq:moment2} follow from the
cosmological redshift of the nonrelativistic momenta and entropy
conservation. In particular,
$d\ln x/dt=H/C$, while collisionless momenta scale as $a_{\rm
FRW}^{-1}$, leading to the $(1-2C)Q$ contribution.

The full thermally averaged annihilation rate is tabulated together
with its logarithmic derivative. Beyond the upper end of the tabulated
$x$ range, the rate is continued using the open-channel $p$-wave
scaling when $T_\chi<T$. The thermodynamic functions are not
extrapolated beyond their tabulated range. For a vanishing initial
abundance we set $(Y,Q)=(0,0)$, so that no kinetic temperature need be
assigned before the dark matter population is produced.
\section{Zero-energy potential bound}
\label{app:potential}

For the potential introduced in Sec.~\ref{wrg}, we define
\begin{equation}
W(R)=m_\chi |V_\chi(R)|
\end{equation}
and write the regular zero-energy radial solution as
$u_\ell(R)=R^{\ell+1}f_\ell(R)$. The radial equation may then be written as
\begin{align}
f_\ell(R)
&=
1-\frac{1}{2\ell+1}
\int_0^R dR'\,R'W(R')
\nonumber\\
&\quad\times
\left[
1-\left(\frac{R'}{R}\right)^{2\ell+1}
\right]
f_\ell(R').
\end{align}

For the two-Yukawa potential,
\begin{equation}
\zeta
=
\int_0^\infty dR\,R W(R)
=
m_\chi
\left(
\frac{\alpha_1}{m_1}
+
\frac{\alpha_2}{m_2}
\right).
\end{equation}
The bound-state counting estimate gives
\begin{equation}
N_\ell<\frac{\zeta}{2\ell+1},
\end{equation}
so that $\zeta<1$ excludes potential bound states.

For the same range, the positivity of the kernel implies
$0<f_\ell(R)\leq1$ and
\begin{equation}
A_\ell\geq1-\frac{\zeta}{2\ell+1},
\end{equation}
where $A_\ell$ is the asymptotic normalization of the zero-energy
solution. The corresponding $p$-wave enhancement therefore satisfies
\begin{equation}
S_1(0)=A_1^{-2}
\leq
\left(1-\frac{\zeta}{3}\right)^{-2}.
\end{equation}

At finite velocity, $S_1$ is obtained by solving the radial equation
numerically and matching the solution to free spherical waves. For the
low-velocity convergence test, the integration endpoint in units of
$m_2R$ is increased from
$40/\min(1,m_1/m_2)$ to $60/\min(1,m_1/m_2)$.
The bounds above refer only to the nonrelativistic exchange potential;
hard radiative corrections are not included.

\section{Independent collision and distribution checks}\label{app:checks}
For a scalar bath with equilibrium distribution $f_b$, we define the
pressure transfer per dark matter particle as
\begin{align}
\mathcal H_P(T_\chi,T)
&=
\int d\mathcal P_\chi
\int\frac{d^3k}{(2\pi)^3} f_b(k)\,v_{\rm M}
\nonumber\\
&\quad\times
\int d\sigma
\left(
\frac{p'^2}{3E'}-\frac{p^2}{3E}
\right),
\end{align}
where $d\mathcal P_\chi$ is the normalized Maxwell--Boltzmann
momentum measure and $v_{\rm M}$ is the M{\o}ller velocity. The
incoming kinematics are treated relativistically, and the differential
cross section is obtained from the full scalar elastic amplitude
generated with \textsc{CalcHEP}. In the dilute-bath limit the outgoing
azimuth can be integrated analytically. Writing
$E'=a+b\cos\varphi$, one finds
\begin{equation}
\left\langle \frac{1}{E'} \right\rangle_\varphi
=
\frac{1}{\sqrt{a^2-b^2}},
\end{equation}
so that the pressure transfer can be evaluated without expanding in
the momentum transfer.

For $T_\chi\neq T$, we define
\begin{equation}
\gamma_P(T_\chi,T)
=
\frac{\mathcal H_P(T_\chi,T)}
     {2(T-T_\chi)}
\end{equation}
and use $\mathcal H_P$ directly in the temperature equation. The
tabulation covers $20\leq x\leq200$ and
$T_\chi/T=0.1,\,0.5,\,0.9,\,1.1$. Increasing the Gaussian quadrature
order from 24 to 36 changes the pressure-transfer rates by less than
$1.1\times10^{-4}$ in relative terms. At $T_\chi=T$, the residual
pressure transfer is below $2.3\times10^{-7}$ when normalized to
$T$ times the collision rate.

For the momentum-resolved calculation, the angular-averaged
annihilation kernel is written as
\begin{align}
K(p,q)
&=
\frac{1}{2}
\int_{-1}^{1} d\mu\,
[\sigma v_{\rm cm}](s)
\frac{s}{4E_pE_q},
\\
s
&=
2m_\chi^2+2(E_pE_q-pq\mu).
\end{align}
The factor $s/(4E_pE_q)$ converts the center-of-mass relative velocity
to the M{\o}ller velocity in the cosmological frame. The cross section
is tabulated from the full $\chi\chi\to h_2h_2$ amplitude, which
accounts for more than $99.99\%$ of the thermal annihilation rate at
the tested temperatures. The inverse process is evaluated with the
same kernel and the equilibrium distribution.

Elastic scattering is described by
\begin{equation}
C_{\rm el}[f]
=
\frac{\gamma(T)}{p^2}
\frac{\partial}{\partial p}
\left[
p^2
\left(
pf+TE_p\frac{\partial f}{\partial p}
\right)
\right].
\end{equation}
Its stationary solution is proportional to $e^{-E_p/T}$. We use
comoving radial momentum to remove the free-streaming term and a
finite-volume discretization that preserves particle number and the
equilibrium solution. The annihilation term is treated implicitly.
The resulting momentum-space evolution remains non-negative in the
cases considered here.
\subsection{Extended-grid numerical controls}\label{app:mapcontrols}
The numerical implementation was checked against independent
\textsc{micrOMEGAs} and \textsc{CalcHEP} calculations. Across the
refitted grid, the single-temperature result agrees with
\textsc{micrOMEGAs} at the sub-permille level, while the scalar
scattering amplitudes agree with \textsc{CalcHEP} to much higher
precision at representative points.

Grid refinement of the annihilation, momentum-transfer and
momentum-space calculations produces changes in the final relic
abundance well below the percent-level kinetic effects discussed in
the main text. The discretized Fokker--Planck operator conserves
particle number and reproduces the equilibrium distribution, and the
FP relic refits reach the target abundance with residuals below
$10^{-4}$.

The finite-transfer scalar-pressure calculation shows similarly small
numerical variations under quadrature refinement. These checks indicate
that the numerical uncertainties are negligible compared with the
kinetic corrections reported in the main analysis.
\section{Low-temperature background check}\label{app:thermal}
The relic-density calculation is performed in the zero-temperature
broken phase. To check the validity of this approximation over the
reheating range considered here, we use the one-loop finite-temperature
effective potential
\begin{equation}
V_{\rm eff}
=
V_0+V_{\rm CW}+V_{\rm ct}+V_T+V_{\rm ring},
\end{equation}
where $V_{\rm CW}$ is the Coleman--Weinberg contribution,
$V_{\rm ct}$ contains counterterms chosen to preserve the
zero-temperature minimum and scalar curvature matrix, and
$V_{\rm ring}$ denotes the Arnold--Espinosa daisy term
\cite{Arnold:1992fb}. The calculation is carried out in Landau gauge
and is used here only to follow the thermal displacement of the scalar
backgrounds.

At the reheating temperature, the shifts of the Higgs and singlet
backgrounds remain small. Without daisy resummation we find
\begin{equation}
\left|\frac{h(T_R)}{v}-1\right|<0.0061,
\qquad
\left|\frac{s(T_R)}{w}-1\right|<10^{-6},
\end{equation}
while inclusion of the Arnold--Espinosa term gives corresponding
maxima of $0.0036$ and $3.5\times10^{-6}$. A one-percent displacement of either background occurs only at
temperatures of about 46~GeV without daisies and 52~GeV with the
Arnold--Espinosa prescription, above the adopted reheating range
$T_R=15$--$40$~GeV.

As a further check, the finite-temperature background values are
inserted into the tree-level parametrization and the relic abundance is
recalculated with the backgrounds held fixed. The resulting change is
below $10^{-4}$ in relative terms. This confirms that the use of the
zero-temperature broken-phase parameters has a negligible effect on the
relic calculation over the temperature range considered here.

The singlet restoration scale obtained from the same potential lies
several hundred GeV above the adopted reheating temperatures. The
high-temperature phase structure is therefore not encountered in the
cosmological history used in this work. A history reaching such
temperatures would also restore the discrete symmetry and reintroduce
the domain-wall problem associated with the degenerate
$s=\pm w$ vacua.
\end{appendices}
\setlength{\bibsep}{0.7em}
\bibliography{references}
\end{document}